\documentclass[aps,prl,twocolumn,superscriptaddress,amsmath,amssymb,10pt]{revtex4-2}

\usepackage{graphicx}
\usepackage{tensor}
\usepackage{xcolor}
\definecolor{linkblue}{RGB}{50,110,175}
\usepackage[bookmarks=true,colorlinks=true,
linkcolor=blue,urlcolor=linkblue,citecolor=blue]{hyperref}
\allowdisplaybreaks
\newcommand{\orcid}[1]{\href{https://orcid.org/#1}{\includegraphics[height=1.9ex,width=1.9ex]{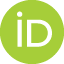}}}
\renewcommand{\vec}{\boldsymbol}
\renewcommand\a{\alpha}
\renewcommand\b{\beta}
\renewcommand\d{\delta}
\renewcommand\k{\kappa}
\renewcommand\l{\lambda}
\renewcommand\r{\rho}

\renewcommand\u{\upsilon}
\renewcommand\c{\chi}

\renewcommand\o{\omega}
\newcommand\e{\epsilon}
\newcommand\g{\gamma}
\newcommand\z{\zeta}
\newcommand\m{\mu}
\newcommand\n{\nu}

\newcommand\p{\pi}
\newcommand\h{\theta}
\newcommand\s{\sigma}

\newcommand\w{\eta}
\newcommand{\ve}{\varepsilon}
\renewcommand\L{\Lambda}
\renewcommand\P{\Pi}
\renewcommand\S{\Sigma}
\renewcommand\O{\Omega}
\renewcommand\H{\Theta}
\newcommand\D{\Delta}
\newcommand\F{\Phi}
\newcommand{\Y}{\Upsilon}
\newcommand{\fig}[1]{Fig.~\ref{#1}}
\newcommand{\eq}[1]{Eq.~(\ref{#1})}
\newcommand{\eqs}[2]{Eqs.~(\ref{#1})-(\ref{#2})}
\newcommand{\txteq}{{\text{eq}}}
\newcommand\lb{\left(}
\newcommand\rb{\right)}
\newcommand\ls{\left[}
\newcommand\rs{\right]}

\newcommand{\lan}{\langle}
\newcommand{\ran}{\rangle}
\newcommand{\non}{\nonumber\\}
\newcommand\pt{\partial}
\newcommand{\nb}{{\nabla}}
\newcommand{\eg}{\emph{e.g.}}
\newcommand{\cO}{{\mathcal{O}}}
\newcommand{\tv}{{\mathrm{v}}}
\newcommand{\bpt}{{\vec{\pt}}}
\newcommand{\bo}{{\vec{\o}}}
\newcommand{\bv}{{\vec{\mathrm{v}}}}
\newcommand{\bx}{{\vec x}}
\newcommand{\bk}{{\vec k}}

\begin{document}

\title{Hydrodynamics of dilation and spin currents}

\author{Zhong-Hua Zhang \orcid{0009-0008-4235-5834}\,}
\affiliation{Physics Department and Center for Field Theory and Particle Physics, Fudan University, Shanghai 200438, China}
\affiliation{Department of Physics, University of Florence, Via G. Sansone 1, I-50019, Sesto F.no (Firenze), Italy}

\author{Xi-Hu Lv \orcid{0009-0005-6035-2600}\,}
\affiliation{Physics Department and Center for Field Theory and Particle Physics, Fudan University, Shanghai 200438, China}

\author{Xu-Guang Huang \orcid{0000-0001-6293-4843}\,}
\email{huangxuguang@fudan.edu.cn}
\affiliation{Physics Department and Center for Field Theory and Particle Physics, Fudan University, Shanghai 200438, China}
\affiliation{Key Laboratory of Nuclear Physics and Ion-beam Application (MOE), Fudan University, Shanghai 200433, China}
\affiliation{Shanghai Research Center for Theoretical Nuclear Physics, National Natural Science Foundation of China and Fudan University, Shanghai 200438, China}

\begin{abstract}
We formulate a relativistic hydrodynamic theory for fluids with spin and intrinsic dilation charges. Using an entropy-current analysis, we derive constitutive relations featuring a bulk viscosity and a dilation conductivity governing the relaxation and diffusion of dilation charge. Linear mode analysis reveals a gapped dilation excitation and the freeze-out of long-wavelength sound modes, similar to the superhorizon modes in cosmology. In the nonrelativistic limit, the theory reduces to that of microstretch fluids. Upon coupling to electromagnetic field, we show that the scale anomaly permits additional contributions in the electric current, dilation current, and energy-momentum tensor. Our theory naturally applies to nearly conformal fluids undergoing rapid expansion or contraction.
\end{abstract}

\maketitle

\pdfbookmark[1]{Introduction}{sec-intro}
{\it Introduction.---}Relativistic hydrodynamics provides a universal macroscopic description of many-body systems near local thermodynamic equilibrium, with applications ranging from the quark–gluon plasma (QGP) in heavy-ion collisions to astrophysical and cosmological fluids~\cite{Romatschke:2017ejr,Rezzolla:2013dea,Weinberg:1972kfs}. It governs the long-wavelength dynamics of conserved quantities, organized by spacetime and global symmetries of the underlying theory.

Recent observations of spin polarization of hadrons in heavy-ion collisions~\cite{STAR:2017ckg,STAR:2022fan} demonstrate that spin can influence macroscopic dynamics. This discovery motivates the extension of relativistic hydrodynamics to include intrinsic angular momentum carried by fluid elements, leading to the framework of spin hydrodynamics~\cite{Florkowski:2017ruc,Florkowski:2018fap,Montenegro:2018bcf,Hattori:2019lfp,Li:2020eon,Fukushima:2020ucl,Hongo:2021ona,Gallegos:2021bzp,She:2021lhe,Cao:2022aku,Daher:2022xon,Gallegos:2022jow}.
See Ref.~\cite{Huang:2024ffg} for a recent review.
In this framework, the total angular momentum
\begin{equation}
    J^{\m\r\s} (x) = x^\r \H^{\m\s} - x^\s \H^{\m\r} + \S^{\m\r\s}
\end{equation}
contains both orbital and spin contributions, with $ \H^{\m\r} $ the energy-momentum tensor and $ \S^{\m\r\s} $ the spin tensor.
The conservation of $ J^{\m\r\s} $ provides independent dynamical equations governing the evolution of spin density. As a result, spin hydrodynamics naturally exploits the full Poincar\'e symmetry.

In this Letter, we further extend the hydrodynamic symmetry to the Weyl-Poincar\'e group. Extension to the full conformal group will be explored elsewhere. We consider a dilation(scale)-invariant quantum field theory in four-dimensional Minkowski spacetime. The symmetry generators consist of 4 translations, 6 Lorentz, and 1 dilation transformations~\cite{DiFrancesco:1997nk}. Associated Noether currents include $\H^{\m\n}$, $J^{\m\r\s}$, and dilation (scale or virial) current $S^\m$, with~\cite{Mack:1969rr,Polchinski:1987dy}
\begin{equation}
\label{def-conformal_currs}
    S^{\m} (x) = \H^{\m\r} x_\r+\Y^{\m} \,,
\end{equation}
where the intrinsic dilation current $ \Y^\m $ depends on $ x $ only via the underlying fields. This extension introduces an extra conservation (balance) equation, suggesting an extension of spin hydrodynamics to include intrinsic dilation as independent degrees of freedom. 

The definition of these intrinsic local currents, however, depends on the pseudo-gauge choice (see Supplemental Materials), and different pseudo-gauges are not physically equivalent.
For example, one may remove $\Y^\m$ and $\S^{\m\r\s}$, rendering $ \H^{\m\n} $ symmetric and traceless, by choosing an appropriate pseudo-gauge~\cite{Callan:1970ze,Polchinski:1987dy}. While many studies of dilation (or even conformal)-invariant hydrodynamics adopt such a pseudo-gauge~\cite{Baier:2007ix,Bhattacharyya:2007vjd,Loganayagam:2008is,Erdmenger:2008rm,Gubser:2010ze,Gubser:2010ui,Eling:2013bj}, we intentionally avoid this to preserve spin and intrinsic dilation as dynamical variables.

A related nonrelativistic framework has appeared in the theory of microstretch fluids~\cite{Eringen:1969a,Eringen:book2}.
In the nonrelativistic limit, spin hydrodynamics reduces to the theory of micropolar fluids~\cite{Eringen:book2,Eringen:1966a,Stokes:1984a}, where microrotation of fluid elements plays the role of spin.
By introducing an additional ``stretch” degree of freedom, one obtains a micropolar fluid with stretch, in which microrotation and microstretch jointly govern the non-Newtonian response of the fluid~\cite{Eringen:1969a,Eringen:book2}.
In a dilation-invariant fluid, the dilation generator (charge) provides a natural relativistic analogue of such a microstretch degree of freedom, thereby suggesting a spin-dilation fluid.

In this Letter, we develop a relativistic hydrodynamic framework that incorporates the spin and intrinsic dilation currents. We derive constitutive relations for these currents and identify new transport coefficients.
The linear mode analysis uncovers a new collective mode associated with the intrinsic dilation. Such a mode naturally arises in nearly scale-invariant fluids undergoing rapid expansion or compression, including high-temperature QGP created in ultra-relativistic heavy-ion collisions and the radiation-dominated early universe.

\pdfbookmark[1]{Phenomenological derivation}{sec-derivation}
{\it Phenomenological derivation.---}Hydrodynamic equations can be constructed phenomenologically from conserved currents, with constitutive relations determined by local thermodynamics and a power counting scheme~\cite{Landau:1987wop}.
The conservation laws of the currents read
\begin{subequations}\label{eq-cons}
\begin{align}
    \label{eq-cons0}
    \pt_\m \H^{\m\n} &= 0 \,,\\
    \label{eq-cons1}
    \pt_\m\S^{\m\r\s} &= \H^{\s\r}-\H^{\r\s} \,,\\
    \label{eq-cons2}
    \pt_\m \Y^\m	&= -\H\indices{^{\m}_\m} \,.
\end{align}
\end{subequations}
Our purpose is to establish a quasi-hydrodynamic theory for spin and dilation based on Eqs.~\eqref{eq-cons}.

We consider a neutral fluid characterized by the local variables: the energy density $ \ve $, the fluid four-velocity $ u^\m $, the spin density $ \s^{\m} $, and the dilation density $ \u $.
We work in the Landau-Lifshitz frame, where the velocity aligns with the energy current and satisfies $ \H^{(\m\n)} u_\n = -\ve u^{\m} $.
We adopt the shorthand notations $ T^{(\m\n)} = (T^{\m\n}+T^{\n\m})/2 $ and $ T^{[\m\n]} = (T^{\m\n}-T^{\n\m})/2 $.
Using the mostly positive Minkowski metric $ \w_{\m\n} = \text{diag}(-,+,+,+) $, the time-like fluid velocity obeys $ u\!\cdot\! u = -1 $.
We take $ \S^{\m\r\s} $ to be totally antisymmetric, following Ref.~\cite{Hongo:2021ona}.
The tensor-form spin density $ \s_{\r\s} = -u^\m \S_{\m\r\s} $ then satisfies $ \s_{\r\s} u^\s = 0 $, leaving three rotational degrees of freedom.
They can be equivalently represented by the dual spin density vector $ \s^\m = \frac{1}{2}\e^{\m\n\r\s}u_\n\s_{\r\s} $.
Finally, we define the dilation density as $ \u = - \Y^\m u_\m $.
We assume the entropy density $ s(\ve, \s^\m, \u ) $ satisfies the local first law
\begin{equation}
    Tds = d\ve - \o_\m d\s^\m - \l d\u \,,
\end{equation}
where $ T $ is temperature, $ \o_\m $ is the spin potential with $ \o\!\cdot\! u = 0 $~\cite{Hongo:2021ona,Cao:2022aku}, and $ \l = -T \pt s/\pt\u $ is the dilation potential conjugate to $ \u $.

Hydrodynamics can be viewed as a low-frequency, long-wavelength effective theory of the underlying field theory.
We thus construct the constitutive relations order by order in derivatives.
To describe fluids that undergo strong expansion or contraction, we adopt a special power-counting scheme in which the scalar expansion $ \h = \pt_\m u^\m $ is treated as a zeroth-order quantity, while other spatial derivatives of velocity remain lower-order, as in standard hydrodynamics.
This power counting is justified because the conformal Killing equation
\begin{equation}\label{eq-killing}
    \pt_\m\b_\n + \pt_\n\b_\m = \frac{1}{2} \w_{\m\n} \,\pt\!\cdot\!\b 
\end{equation}
allows arbitrary $\h$ to exist at global equilibrium, where $ \b_\m = \b u_\m $ denotes the thermal current and $ \b = 1/T $ the inverse temperature.
At global equilibrium, $ \h = 3\l $.
So, we regard both $ \l $ and $ \u $ as zeroth-order quantities, while $ \h-3\l $ counts as $ \cO (\pt) $.
The spin potential scales as $\cO(\pt) $ and the spin density as $ \cO(\hbar\pt) $, reflecting the quantum nature of spin.
At leading order, SO(3) symmetry is preserved, allowing us to decompose the currents using $ u^\m $ and $ \w^{\m\n} $ as the zeroth-order quantities
\begin{subequations}\label{constitutive}
\begin{align}
    \H^{\m\n} =&~ \ve u^\m u^\n + p\D^{\m\n} +\d\H^{\m\n}\,,\\
    \S^{\m\r\s} =&~ \e^{\m\r\s\n} \lb \s_\n + u_\n \d\s \rb \,,\\
    \Y^\m =&~ \u u^\m + \d\Y^\m \,,
\end{align}
\end{subequations}
where $ \D^{\m\n} = \w^{\m\n} + u^\m u^\n $ denotes the transverse projection.
One may further decompose the correction
\[
\d\H^{\m\n} = \P\D^{\m\n} +\d\H^{\m\n}_s+ u^{\m} q^{\n} -u^{\n} q^{\m} + \d\H^{\m\n}_a \,,
\]
where the tensor $ \d\H_{s}^{\m\n} $ is traceless and symmetric, while $ \d\H^{\m\n}_a $ is antisymmetric.
The corrections obey $ \d\H^{\m\n}_s u_\n = \d\H^{\m\n}_a u_\n = q^\m u_\m = \d\Y^{\m} u_\m = 0 $.
We distinguish the static pressure $ p $ from the viscous pressure $ \P $, the latter containing purely dissipative contributions.
Since $ \S^{\m\r\s} $ is totally antisymmetric, relation $ u_\r\pt_\m\S^{\m\r\s} = -2u_\r\H^{[\r\s]} $ imposes constraints rather than providing dynamical evolution. This implies $ q^\m = -\frac{1}{2} \e^{\m\n\r\l}u_\n\pt_\r \s_\l \sim  \cO(\hbar\pt^2) $, which is higher-order in gradients and will be neglected in the following. Furthermore, one can show that $\d\s\sim \cO(\hbar\pt^2)$ and thus is also negligible~\cite{Hongo:2021ona}. 

To illustrate the entropy analysis explicitly, we perform a Legendre transformation and define the dimensionless grand potential density \( f = -s + \b\ve -\O^\m\s_\m  -\L\u \)
with $ \L = \b\l $ and $ \O_\m = \b\o_\m $ the reduced dilation and spin potentials.
From the definition of $ f $, we express the entropy current as~\cite{Israel:1979wp,Becattini:2023ouz}
\begin{equation}
    s^\m = -f u^\m -\H^{\m\n}\b_\n -\frac{1}{2}\S^{\m\r\s} \O_{\r\s} - \Y^\m\L +\d s^\m
\end{equation}
with $ \O_{\r\s} = -\e_{\r\s\m\n} u^\m \O^\n $ the dual tensor of reduced spin potential and $ \d s^\m $ standing for the intrinsic ambiguity in defining $ s^\m $.
Using the conservation (balance) equations~\eqref{eq-cons} and the chain rule
\(
    \pt_\m f = \ve \pt_\m\b - \s_\n\pt_\m\O^\n - \u\pt_\m\L \,,
\)
we derive the entropy production rate
\begin{equation}\label{entropy-production}
\begin{split}
    \pt_\m s^\m =& -\h(f+\b p) +\b\l \lb 3p - \ve \rb -\b\P(\h-3\l) \\
    &-\b\d\H^{\m\n}_s \nb_{\lan \m} u_{\n\ran} - \b\d\H^{\m\n}_a (\nb_{[\m} u_{\n]}-\o_{\m\n})\\
    &- \d\Y^\m \nb_\m \L + \pt_\m\d s^\m +\cO(\pt^3,\hbar\pt^2)
\end{split}
\end{equation}
with $ \nb_\m = \D_{\m\n}\pt^\n $ the spatial derivatives, $ \nb_{\lan \m} u_{\n\ran} = \nb_{(\m} u_{\n)}-\h\D_{\m\n}/3 $ the shear tensor, and $\o_{\m\n}=T \O_{\m\n}$.
Defining the static pressure by $ p = -Tf $, one finds that $ \ve - 3p = 0 $ for fluids that preserve dilation symmetry. Thus, the first two terms in \eq{entropy-production} vanish.
The second law of thermodynamics is then satisfied by setting $ \d s^\m = 0 $ and identifying the dissipative corrections as
\begin{align}\label{constitutive-relations}
    \d\H_s^{\m\n} &= -2\w \nb^{\lan \m} u^{\n\ran} \,, \quad
    \P = -\z (\h-3\l) \,,\\
    \d\H_a^{\m\n} &= -2\w_s \lb\nb^{[\m} u^{\n]}-\o^{\m\n}\rb \,, \,\,
    \d\Y^\m = -\k_d T \nb^\m\L \,, \nonumber
\end{align}
where $ \w $, $ \z $, $ \w_s $, and $ \k_d $ are nonnegative transport coefficients.
Substituting constitutive relations \eqref{constitutive-relations} into Eqs.~\eqref{constitutive}, and applying conservation laws \eqref{eq-cons}, one obtains the full set of relativistic hydrodynamic equations.

In Eqs.~\eqref{constitutive-relations}, $ \w $ is the shear viscosity, while $ \w_s $, known as the rotational viscosity~\cite{Hattori:2019lfp}, characterizes spin relaxation.
The bulk viscosity $ \z $ plays a role distinct from the conventional one. In a scale-invariant fluid, resistance to uniform expansion and compression is absent, implying a vanishing conventional bulk viscosity.
In contrast, here $\z$ can be finite, which measures the relaxation rate of intrinsic expansion of the fluid cell, determined by $\l$, toward the equilibrium value set by the macroscopic expansion $ \h /3 $.
This effect becomes relevant whenever rapid stretching or compression of a fluid excites sizable intrinsic dilation.
Meanwhile, the coefficient $ \k_d $ controls diffusion of the dilation density driven by the gradient of $ \l/T $, which motivates the name {\it dilation conductivity}.

\pdfbookmark[1]{Linear mode analysis}{sec-linear}
{\it Linear mode analysis.---}We linearize the hydrodynamic equations around a global equilibrium configuration and analyze the resulting eigenmodes near the spacetime origin.
We adopt a thermal current
\(
    \b^\m_\txteq(x) = \lb t^\m + \l_0 x^\m\rb/T_0
\),
which solves the conformal Killing equation~\eqref{eq-killing} and describes a global equilibrium with uniform expansion or contraction.
Here $ t^\m = (1,\vec{0}) $, while $ T_0 $ and $ \l_0 $ are constants.
We restrict the analysis to region $ (\l_0 \bx )^2 < (1 + \l_0 t)^2 $, in order to ensure that $ \b^\m $ remains timelike.
We further assume that $ |\l_0| $ is small so that the equilibrium region is sufficiently large for linear analysis. 
Equilibrium temperature and velocity are $ T_{\txteq}(x) = T_0 / \D(x) $ and $ u_{\txteq}^\m(x) = (t^\m + \l_0 x^\m) / \D(x) $, with $ \D(x) = \sqrt{(1 + \l_0 t)^2 - (\l_0 \bx )^2 }$.
At global equilibrium, the dissipative corrections~\eqref{constitutive-relations} vanish, implying an equilibrium dilation potential $ \l_\txteq(x) = \l_0/\D(x) $ and a vanishing equilibrium spin potential.
The ideal hydrodynamic equations then yield the equilibrium densities $ \ve_\txteq(x) = \ve_0 \D(x)^{-4} $, $ \u_\txteq(x) = \u_0 \D(x)^{-3} $, and $ \s^\m_\txteq(x) = 0 $.
Here, the subscript ``eq" marks equilibrium quantities, while ``0" denotes their values at the origin.
Consider linear plane-wave perturbations of densities, potentials, and velocity around this equilibrium background.
They are marked by $ \d $, \eg, $ u^\m (x) = u_\txteq^\m(x) + \d u^\m (x) $.

Substituting perturbations into equations of motion, we linearize them around the spacetime origin as
\begin{subequations}\label{EOM-linear}
\begin{align}
\label{EOM-linear-e}
   0 =& \pt_0 \d\ve + \lb 1 - 3\l_0 \z' \rb \pt_i \d\p^i + \l_0 \lb 4 \d\ve + 3 D_d \d\u \rb \,, \\
\label{EOM-linear-v}
    0 =& \pt_0 \d\p^i + c_s^2 \pt^i \d\ve -(\w'+\w_s') (\d^{ij}\bpt^2 -\pt^i\pt^j)\d\p_j \non
    & - \g_\parallel \pt^i\pt^j \d\p_j - D_s \e^{ijk} \pt_j \d\s_{k} + D_d \pt^i \d\u \non
    & + {\l_0}^2 \g_{-} \d\p^i - \l_0 \g_{+} \pt_0 \d\p^i \,, \\
\label{EOM-linear-s}
    0 =& \pt_0 \d\s^{i} + 2 D_s \d\s^{i} - 2\w_s' \e^{ijk} \pt_j \d\p_k \,, \\
\label{EOM-linear-up}
    0 =& \pt_0 \d\u + \lb \u_0^\prime - 3 \z' \rb \pt_i \d\p^i + 3 D_d \d\u - \k_d^\prime \bpt^2 \d\u \non
    & + C_{\u} \bpt^2 \d\ve + 3 \l_0 \lb \d\u - \k_d^\prime \pt_0 \d\u + C_{\u} \pt_0 \d\ve \rb \,,
\end{align}
\end{subequations}
where $ \d\p_i = (\ve_0+p_0) \d u_i $, $ \bpt^2 = \pt_k \pt^k $, and repeated spatial indices are summed.
We have introduced the spacetime constants
\[
\begin{gathered}
    c_s^2 = \frac{\pt p}{\pt\ve}=\frac{1}{3} \,,\quad
    \u_0^\prime     = \frac{\u_0}{\ve_0 + p_0} \,, \quad
    \w^\prime  = \frac{\w}{\ve_0+p_0} \,,\\
    \z^\prime  = \frac{\z}{\ve_0+p_0} \,,\quad
    \w_{s}^\prime  = \frac{\w_s}{\ve_0+p_0} \,,\quad
    \g_\parallel    = \z^\prime+\frac{4}{3}\w^\prime \,,\\
    \g_- = \g_\parallel + 2\w' - 2\w_s' \,,\quad
    \g_+ = \g_\parallel + 2\w' + 2\w_s' \,,\\
    \c_d  = \frac{\pt\u}{\pt\l} \,,\quad
    D_d   = \frac{3\z}{\c_d} \,, \quad
    \c_s            = \frac{\pt \s^{i}}{\pt\o^{i}} \,, \\
    D_s             = \frac{2\w_s}{\c_s} \,,\quad
    \k_d^\prime     = \frac{\k_d}{\c_d} \frac{\ve_0+p_0}{T_0s_0}\,,\quad
    C_{\u}       = \frac{\k_d\l_0}{3T_0s_0} \,.
\end{gathered}
\]
In momentum space, where $ \pt_0 \to -i \o $ and $ \pt_i \to i k_i $, the differential equations reduce to the algebraic equation $ M(\o,\bk) \d c(\o, \bk) = 0 $, where $ \d c $ collects the dynamic variables. Choosing $ \bk = (0,0,k) $ without loss of generality, $ M $ becomes block diagonal with independent sectors $ \{\d\s_3\} $, $ \{\d\p_1, \d\s_2\} $, $ \{\d\p_2, \d\s_1\} $, and $ \{\d\ve,\d\p_3, \d\u\} $. The eigenmodes of each sector follow from $ \det M(\o,k) = 0 $.

\begin{figure}[!t]
    \centering  
    \begin{minipage}[b]{0.235\textwidth}
        \includegraphics[width=\linewidth]{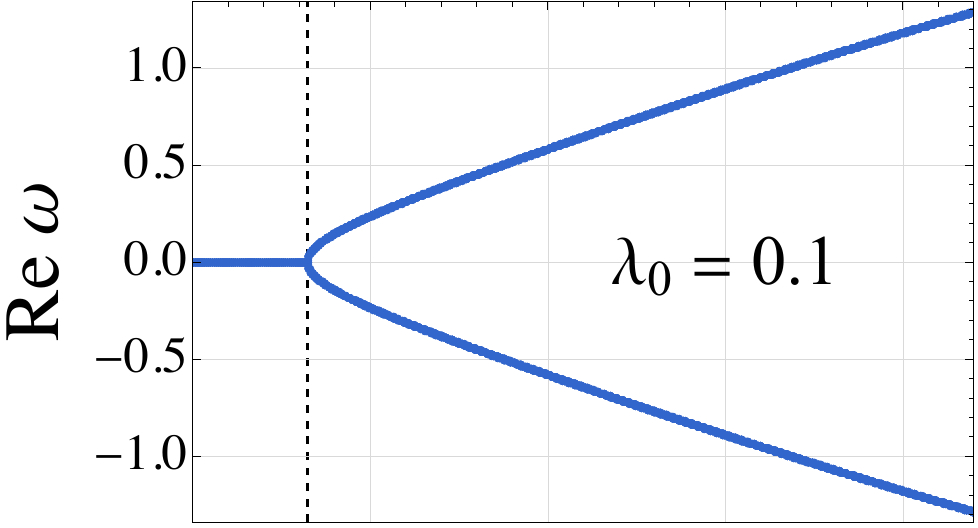}
    \end{minipage}
    \begin{minipage}[b]{0.235\textwidth}
        \includegraphics[width=\linewidth]{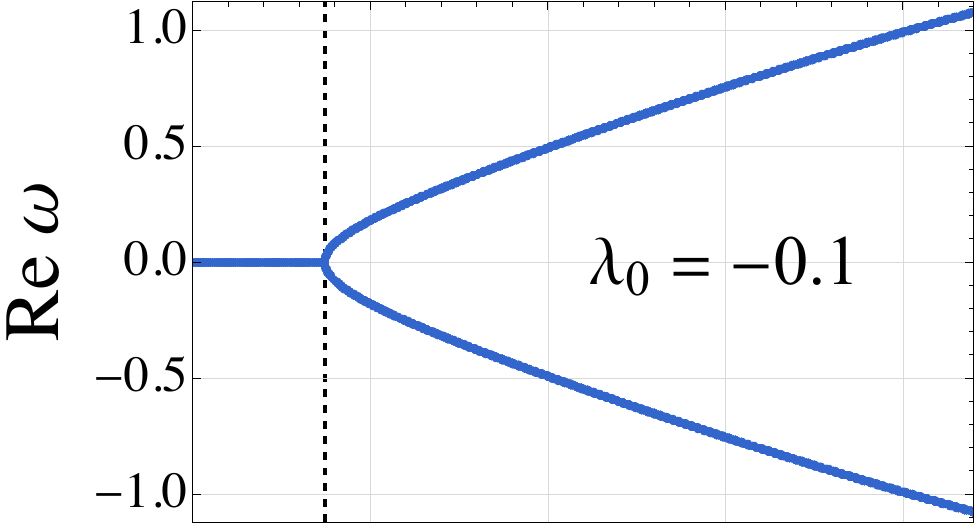}
    \end{minipage}
    
    \vspace{-0.06cm}
    
    \begin{minipage}[b]{0.235\textwidth}
        \includegraphics[width=\linewidth]{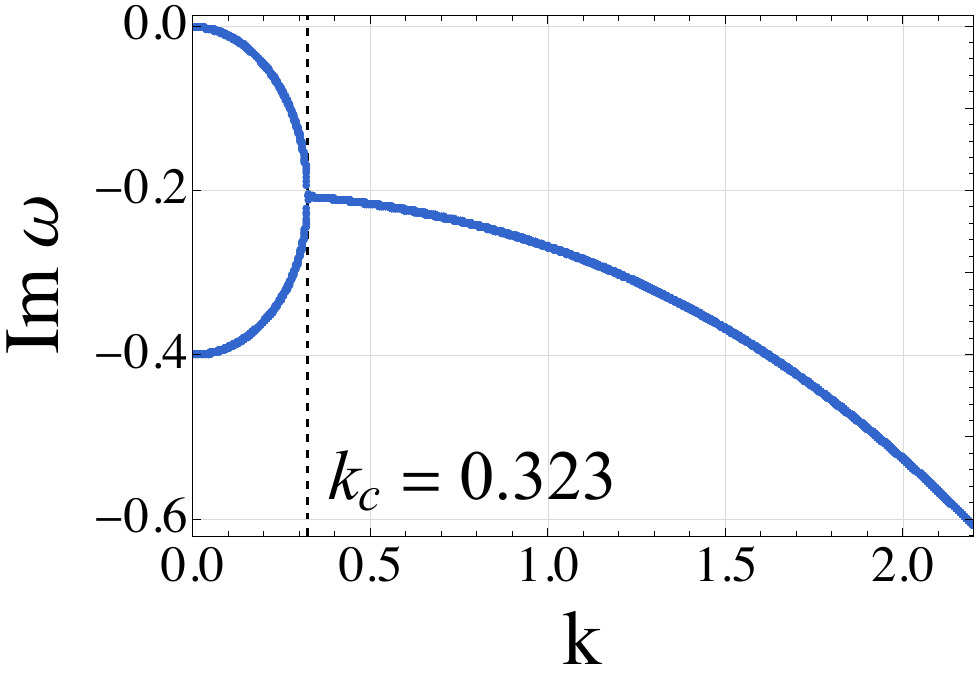}
    \end{minipage}
    \begin{minipage}[b]{0.235\textwidth}
        \includegraphics[width=\linewidth]{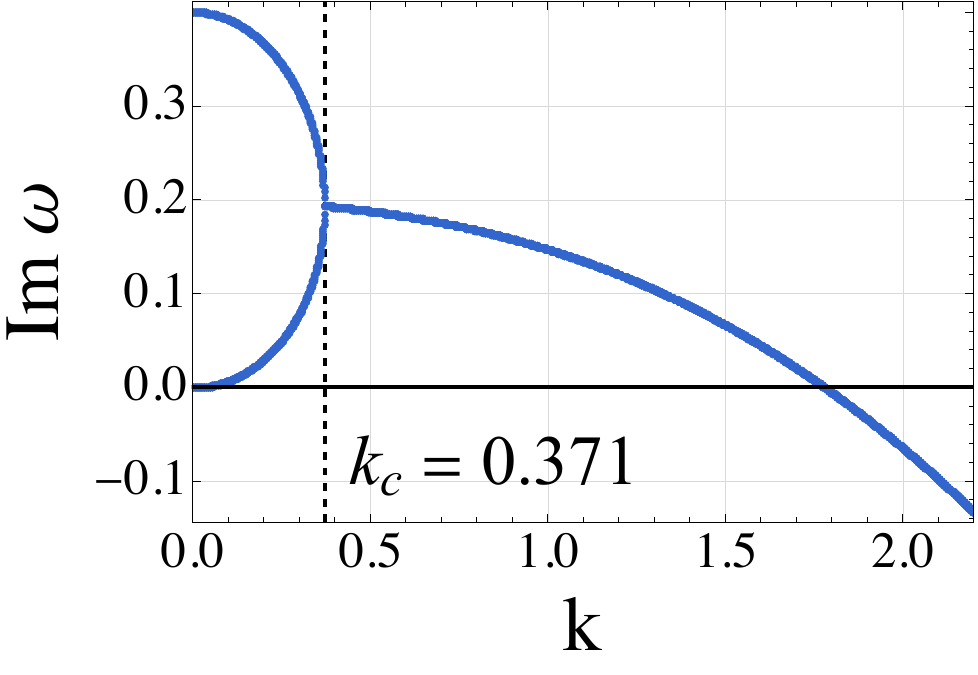}
    \end{minipage}
    \caption{The left (right) column exhibits the sound modes' dispersion relation of dilation-invariant hydrodynamics with a background expansion (contraction).
    We use units with $ T_0 = 1 $.
    So, all physical quantities are dimensionless.
    We take $ p_0 = 16 \p^2/90 $, $ \ve_0 = 3 p_0 $, $ \w = s/(4\p)$, and $ \z = \w_s = \k_d = \c_d = 1 $.
    We also assume that $ \u_0 = \c_d \l_0 $, which is justified when $ \l_0 $ is small.}
    \label{fig-dispersion}
\end{figure}

Perturbation $ \d\s_3 $ yields a purely relaxational longitudinal spin mode $ \o = -2iD_s $~\cite{Hattori:2019lfp,Hongo:2021ona}.
Coupled perturbations $ \{\d\p_1, \d\s_2\} $ and $ \{\d\p_2, \d\s_1\} $ give two shear modes (gapless at the order considered) and two gapped spin modes
\begin{equation}
    \o =
    \begin{cases} \displaystyle
        -i\w'k^2 \lb 1 + \g_{+}\l_0 \rb +\cO(\l_0^2,k^3)   \,, \\
  -2i D_s -i\w_s' k^2 \lb 1+\g_{+}\l_0\rb +\cO(\l_0^2,k^3)     \,. 
    \end{cases}
\end{equation}
Note that shear modes exhibit a small gap at order $ \l_0^2 $. The remaining sector $ \{\d\ve,\d\p_3,\d\u\} $ gives one sound-mode pair and one gapped dilation mode. The dilational dispersion reads
\begin{equation}
\begin{split}
    \o =& -3i D_d (1+3\k_d^\prime\l_0) -3i\l_0\\
    & - i(\z'+\k_d^\prime+a)k^2 + \cO(\l_0^2,k^3) \,,
\end{split}
\end{equation}
where
\(
    a = \z'\g_{+}\l_0 + 3(\k_d^\prime)^{2}\l_0 -\frac{(1-c_s^2)\c_d\l_0+\u_0}{3(\ve_0+p_0)}
\).
This mode damps with a characteristic lifetime $ \approx1/(3D_d) $.
Sound-mode dispersions are shown in \fig{fig-dispersion}.
In the zero-wavelength limit, the sound modes behave as $ \o = -i\l_0^2 \g_{-} $  and $ \o = -4i\l_0 $ to leading order in $ \l_0 $, corresponding to momentum and energy dilution or concentration due to the expansion or contraction.
Moreover, long-wavelength perturbations freeze, which means that modes with $ k\leq k_c $ do not propagate.
The critical momentum $k_c$ follows from the vanishing discriminant of the cubic equation in $ i\o $ in the $ \{\d\ve,\d\p_3,\d\u\} $ sector,
\begin{equation}
    k_c = \frac{2 |\l_0|}{c_s} \ls 1- \l_0 \g_c +\cO(\l_0^2) \rs
\end{equation}
with $ \g_c = \lb \frac{3}{4}\z'+\frac{13}{2}\w'+\frac{1}{2}\w_s'\rb $. The leading term admits an interpretation analogous to the freeze-out of superhorizon modes in cosmological inflation~\cite{Mukhanov:1990me,Dodelson:2003ft}.
In an expanding background, $ \l_0 $ plays the role of a Hubble-like constant, with the horizon radius $ R \sim c_s/ |\l_0| $.
Perturbations with wavelengths larger than $ R $ freeze out because signals cannot propagate across the horizon.
A similar argument applies to contraction, where perturbations outside the horizon cannot reach the origin before collapse.
%

\pdfbookmark[1]{Nonrelativistic limit}{sec-nonrela}
{\it Nonrelativistic limit.---}We obtain the nonrelativistic limit of the equations of motion by taking the four-velocity as $  u^\m \rightarrow (1,\bv) $.
We also treat $ \l $, $ \o_\m $, and $ \nb_\m T $ as $ \cO(\tv) $ quantities, consistent with their behavior at global equilibrium.
Neglecting the $ \cO(\tv^2) $ terms, we find
\begin{subequations}\label{eq-nr}
\begin{align}
\label{eq-nr1}
    0 =& \dot{\ve} + (\ve + p) \pt_i \tv^i \,, \\
\label{eq-nr2}
    0 =& (\ve+p) \dot{\tv}^i + \pt^i p - \lb\z-\w_s+\frac{\w}{3} \rb \pt^i \pt_k \tv^k \non
    & - (\w+\w_s)\bpt^2 \tv^i + 3 \z \pt^i \l -2\w_s\e^{ijk}\pt_j\o_k \,, \\
\label{eq-nr3}
    0 =& \dot{\s}^i -2\w_s\e^{ijk}\pt_j \tv_k + 4\w_s \o^i\,, \\
\label{eq-nr4}
    0 =& \dot{\u} + \u \pt_i \tv^i +3 \z \lb 3\l - \pt_i \tv^i \rb -\k_d \bpt^2 \l \,,
\end{align}
\end{subequations}
where the dot denotes $ \pt_0 + \tv^i \pt_i $.
Note that a strict Newtonian limit does not exist for scale-invariant fluids because the sound speed is comparable to the speed of light. Therefore, the first equation does not recover the continuity equation for mass density. 
Comparing Eqs.~\eqref{eq-nr} with those of microstretch fluids~\cite{Eringen:book2}, we identify $ \l $ with the stretching degree of freedom, and $\o^i$ with the microrotation degree of freedom. Within their linear approximation and our power-counting scheme, Eqs.~\eqref{eq-nr2}-\eqref{eq-nr4} coincide with the microstretch-fluid equations.
Hence, the intrinsic dilation plays the same role as the microstretch of fluid cells, just as spin corresponds to microrotation (see Supplemental Materials for details). 

\pdfbookmark[1]{Scale anomaly}{sec-anomaly}
{\it Scale anomaly.---}So far, we have focused on a neutral fluid. For a charged fluid coupled to an electromagnetic field, scale invariance is broken at the quantum level. Consequently, a scale anomaly term $-C_T F_{\m\n}F^{\m\n}$ must be added to the right-hand side of the balance equation~\eqref{eq-cons2}, where $C_T = \b(e)/2e$ and $F^{\m\n}\sim\cO(\pt)$ is the field strength tensor. Defining the electric current as $j^\m = n u^\m + \d j^\m$, where $n$ is the charge density, the right-hand side of Eq.~\eqref{eq-cons0} becomes $F^{\n\l}j_\l$. The first law of thermodynamics generalizes to $Tds = d\ve - \o_\m d\s^\m - \l d\u - \m dn$, with chemical potential $\m$. Correspondingly, the entropy current becomes
\begin{equation}
	s^\m = p \b^\m -\H^{\m\n}\b_\n -\frac{1}{2}\S^{\m\r\s} \O_{\r\s} - \Y^\m\L - j^\m \a +\d s^\m \,,
\end{equation}
with reduced chemical potential $\a = \b\m$. Direct calculation shows that the entropy production $\pt_\m s^\m$ acquires two additional terms compared to Eq.~\eqref{entropy-production}: $C_T\L F_{\m\n}F^{\m\n} - \d j^\m\lb \pt_\m\a -\b E_\m\rb$, where $E_\m = F_{\m\n} u^\n$ is the rest-frame electric field. The semi-positivity of $\pt_\m s^\m$ requires the dissipative constitutive relation $\d j^\m = -\k \lb T\nb^\m\a - E^\m\rb$, with electrical conductivity $\k\geq 0$.

However, the $C_T\L F_{\m\n}F^{\m\n}$ term is not sign-definite. Furthermore, the scale anomaly induces an additional charge transport~\cite{Chernodub:2016lbo,Chernodub:2017jcp}, 
\begin{align}
\label{scadja}
    \d j_A^\m = 4C_T \pt_\n(F^{\m\n}\L) \,,
\end{align}
which contributes a term $-\d j_A^\m \lb \pt_\m\a -\b E_\m\rb$ that threatens the semi-positivity of entropy production at $\cO(\pt^2)$. To resolve this, we follow the strategy of Son and Sur\'{o}wka for chiral anomalies~\cite{Son:2009tf} by introducing scale-anomaly-induced corrections $\d\H^{\m\n}_A$, $\d\Upsilon^\m_A$, $\d\S^{\m\r\s}_A$, and $\d s^\m_A$. We can safely set $\d\S^{\m\r\s}_A$ to zero, since the spin sector is unaffected by the scale anomaly.
The entropy divergence then splits into normal and anomalous parts, $\pt_\m s^\m = \pt_\m s^\m|_{\rm nor} + \pt_\m s^\m|_{\rm anom}$, where
\begin{align}
\label{entrop2}
	\pt_\m s^\m\big|_{\rm nor} &= \frac{\d\H_s^2}{2\w T} + \frac{\P^2}{\z T} + \frac{\d\H_a^2}{2\w_s T} + \frac{\d\Y^2}{\k_d T} + \frac{\d j^2}{\k T} \,,\\
	\pt_\m s^\m\big|_{\rm anom} &= -\d \H^{\m\n}_A\lb\pt_\m\b_\n-\w_{\m\n}\L -\O_{\m\n}\rb \non
	- \d & j^\m_A \lb\pt_\m\a -\b E_\m\rb -\d\Upsilon_A^\m\pt_\m\L + \pt_\m \d s_A^\m \,.
\end{align}
Here, we have imposed the modified equation of state $ \ve - 3p = C_TF^2 $~\cite{Chernodub:2010sq}, which absorbs $ C_T \L F^2 $.
Substituting $\d j_A^\m$ in Eq.~\eqref{scadja}, we rewrite its contribution as
\begin{align}
	\label{scalere}
	&-\d j_A^\m \lb \pt_\m\a -\b E_\m\rb = -4C_T\L \H^{\m\n}_{\rm EM}\pt_\m\b_\n+C_T F^2\b^\m\pt_\m\L\non
	& +4 C_T\pt_\m\ls\L F^{\m\n}\lb \pt_\n\a -\b E_\n\rb\rs-C_T\pt_\m\lb\L\b^\m F^2\rb \,,
\end{align}
with $\H^{\m\n}_{\rm EM}=F^{\m\l}{F^\n}_\l-F^2\w^{\m\n}/4$ the electromagnetic energy-momentum tensor. Demanding $\pt_\m s^\m\big|_{\rm anom}=0$ to strictly enforce the second law, we identify
\begin{subequations}\label{eq-d}
	\begin{align}
		\label{eq-dsa}
		\d s_A^\m &= C_T\L\b^\m F^2 - 4 C_T\L F^{\m\n}\lb \pt_\n\a -\b E_\n\rb \,, \\
		\label{eq-dHa}
		\d\H^{\m\n}_A &= -4 C_T\L \H^{\m\n}_{\rm EM} \,, \\
		\label{eq-dUa}
		\d\Upsilon_A^\m &= C_T F^2\b^\m \,.
	\end{align}
\end{subequations}
Thus, local thermodynamics dictates that the anomaly-induced current $\d j^\m_A$ must be accompanied by corrections to the energy-momentum tensor and the dilation density $ \d\u_A =-u\cdot\d \Upsilon_A=-C_T F^2/T$. All these anomaly-induced contributions are nondissipative.

\pdfbookmark[1]{Conclusions}{sec-conclusions}
{\it Conclusions.---}We developed a relativistic hydrodynamic theory for scale-invariant fluids with spin and intrinsic dilation via an entropy-current analysis. The resulting constitutive relations \eqref{constitutive-relations} contain a finite viscous pressure that, rather than signaling a scale invariance breaking, drives the relaxation of intrinsic dilation at a rate governed by the {\it bulk viscosity}. Linear mode analysis reveals a gapped dilation mode and a long-wavelength sound-mode freeze-out, analogous to superhorizon freeze-out in cosmological inflation. In the nonrelativistic limit, our framework reduces to those of microstretch fluids, with the intrinsic dilation analogous to microstretch and spin to microrotation, respectively. Furthermore, coupling to an electromagnetic field reveals that the scale anomaly induces novel nondissipative contributions to the electric current, energy-momentum tensor, and dilation density.

Physical systems to which our framework may apply include the strongly expanding quark-gluon plasma produced in relativistic heavy-ion collisions and the expanding, radiation-dominated early universe; for the latter, our Minkowski-space formulation may capture the essential local dynamics, though a complete treatment requires general-relativistic hydrodynamics.

Finally, we note that for nearly scale-invariant fluids, spin may cease to be a good hydrodynamic variable due to strong spin-orbit coupling. One could integrate out the spin density $ \s_\m $ by setting $ \d\H_a^{\m\n} = 0 $ while preserving the remaining constitutive relations~\eqref{constitutive}. Alternatively, one could replace spin with an axial charge $ n_5 $ obeying the anomalous balance equation $\pt_\m j^\m_5 = C_A E\cdot B $, with $C_A$ the chiral-anomaly coefficient. Following Ref.~\cite{Son:2009tf}, this would induce anomalous dilation current $ \Y^\m $ via magnetic fields or vorticity. We leave this possibility for future study and retain spin here to establish a comparison with microstretch fluids.

We acknowledge the helpful discussions with Shuai Wang. Z.-H.Z. also thanks the University of Florence for its hospitality. This work is partly supported by the Natural Science Foundation of Shanghai (Grant No. 23JC1400200), the National Natural Science Foundation of China (Grants No. 12225502 and  No. 12147101), and the National Key Research and Development Program of China (Grant No. 2022YFA1604900). Z.-H.Z. is also supported by the China Scholarship Council (CSC) under Grant No. 202406100158.

\bibliography{biblio.bib}


\onecolumngrid
\clearpage\newpage

\vspace{2em}
\begin{center}
\textbf{\large Supplemental Materials}
\end{center}

\setcounter{section}{0}
\setcounter{figure}{0}
\setcounter{table}{0}
\renewcommand{\thesection}{S\arabic{section}}
\renewcommand{\thefigure}{S\arabic{figure}}
\renewcommand{\thetable}{S\arabic{table}}

\section{Pseudo-gauge transformation}\label{pseudogauge}
One can modify a conserved current by adding another current while preserving the conservation equation and the total conserved charge.
This procedure is called a pseudo-gauge transformation.
For currents $ \H $, $ \S$, and $ \Y $, satisfying balance equations \eqref{eq-cons}, we perform the following transformation
\begin{align}
    \H_T^{\m\n}   &= \H^{\m\n} +\frac{1}{2} \pt_{\l} \lb\F^{\l\m\n} - \F^{\m\l\n} -\F^{\n\l\m}\rb + \pt_\a\pt_\b \Psi^{\m\a\n\b} \,,\\
    \S_T^{\m\r\s} &= \S^{\m\r\s} -\F^{\m\r\s} +\pt_{\l}Z^{\m\l\r\s} \,,\\
    \Y_T^{\m}     &= \Y^{\m} -{{\F^\n}_\n}^\m - \pt_\l \Psi^{\m\a\l}{}_{\a}+\pt_{\l}B^{\m\l} \,,
\end{align}
where $\F^{\m\r\s}$ is antisymmetric on $\r\s$, $Z^{\m\l\r\s}$ and $\Psi^{\m\l\r\s}$ are both antisymmetric on $\m\l$ and $ \r\s $ respectively and symmetric under exchange of $ \m\l $ and $ \r\s $, and $B^{\m\l}$ is antisymmetric on $\m\l$.
It is straightforward to check that balance equations~\eqref{eq-cons} still hold for currents $ \H_T $, $ \S_T$, and $ \Y_T $.

\section{Comparison with microstretch fluid}\label{microstretch}
In this section, we provide additional details on the nonrelativistic limit. In particular, we compare the equations of motion \eqref{eq-nr1}-\eqref{eq-nr4} with those of the microstretch fluid.
After applying the power counting used in this work, Eqs.~(15.4.3)-(15.4.5) in Ref.~\cite{Eringen:book2} read
\begin{align}
\label{sp-stretch-1}
    0 &= \r(\dot{\bv}-\vec{f}) + \bpt \pi - (\lambda_v + \mu_v)\bpt \lb\bpt \cdot \bv\rb - (\mu_v + \kappa_v) \bpt^2 \bv - \lambda_0 \bpt \nu - \kappa_v \bpt \times \vec{\nu} \,, \\
\label{sp-stretch-2}
    0 &= \rho (j\dot{\vec{\n}} -\vec{l}) -\k_\n \bpt \times \bv + 2\k_\n \vec{\n} \,,\\
\label{sp-stretch-3}
    0 &= \rho \lb\frac{j}{2}\dot{\n}-l\rb -\p_0 + \lambda_1 \nu + \lambda_0 \bpt \cdot \bv -\alpha_0 \bpt^{2} \nu \,.
\end{align}
Restricted by the second law of thermodynamics, these transport coefficients satisfy
\begin{equation}\label{entropy-condition}
    3\l_v +2\m_v + \k_v \geq \frac{3\l_0^2}{ \l_1} \,,\quad 2\m_v + \k_v\geq 0 \,,\quad \k_v \,,\, \l_1 \,,\, \a_0 \geq 0 \,.
\end{equation}
Note that we have corrected a typo in Eq.~(15.4.3) of Ref.~\cite{Eringen:book2} by comparing it with Eq.~(5.3) of Ref.~\cite{Eringen:1969a}.
After linearization, replacing $\dot{\vec\sigma}$ and $\dot{\upsilon}$ by $\chi_s \dot{\vec\omega}$ and $\chi_d \dot{\lambda}$ and neglecting the nonlinear term $\upsilon \partial_i \tv^i$, our \eqs{eq-nr2}{eq-nr4} give
\begin{align}
\label{sp-nr-1}
    0 &= (\ve+p) \dot{\bv} + \bpt p - \lb\z-\w_s+\frac{\w}{3} \rb \bpt (\bpt \cdot \bv) - (\w+\w_s) \bpt^2 \bv + 3 \z \bpt \l - 2\w_s \bpt \times \bo \,, \\
\label{sp-nr-2}
    0 &= \c_s \dot{\bo} -2\w_s \bpt\times\bv + 4\w_s \bo \,, \\
\label{sp-nr-3}
    0 &= \c_d \dot{\l} +9 \z\l  - 3\z \bpt\cdot \bv -\k_d \bpt^2 \l \,.
\end{align}
To compare \eqs{sp-nr-1}{sp-nr-3} with \eqs{sp-stretch-1}{sp-stretch-3}, we turn off the external forces $ \vec{f} $, $ \vec{l} $, $ l $, and the inertial micropressure $ \p_0 $ in \eqs{sp-stretch-1}{sp-stretch-3}.
Then, by matching the coefficients in Ref.~\cite{Eringen:book2} to ours as
\begin{equation}
    \r j=\c_s = 2\c_d \,,\quad \l_1 = -3\l_0 = 9\z \,,\quad \l_v = \z-\frac{2\w}{3} \,,\quad \m_v = \w-\w_s \,,\quad \k_v = 2\w_s \,,\quad \a_0 = \k_d \,,
\end{equation}
\eqs{sp-nr-1}{sp-nr-3} reproduce \eqs{sp-stretch-1}{sp-stretch-3} and the conditions \eqref{entropy-condition} are equivalent to $ \z,\w,\w_s,\k_d \geq 0 $. 
Thus, \eqs{sp-nr-1}{sp-nr-3} become identical to \eqs{sp-stretch-1}{sp-stretch-3}, except in the Navier-Stokes equation, where Ref.~\cite{Eringen:book2} uses $\rho\dot{\bv} $ while we use $(\ve+p)\dot{\bv} $.
The evolution of $ \r $ follows from the continuity equation for the mass density,
\begin{equation}
    0 = \dot{\r} + \r \bpt\cdot \bv \,,
\end{equation}
while from \eq{eq-nr1}, $ \ve+ p $ satisfies
\begin{equation}
    0 = (\dot{\ve} + \dot{p}) + \frac{4}{3}(\ve+p) \bpt\cdot \bv \,.
\end{equation}
This difference reflects the fact that a scale-invariant fluid does not possess a genuine Newtonian limit.
Its effective density $ (\ve+p) $ still behaves relativistically at the limit $ \tv\to 0 $.
\end{document}